\documentclass[12pt]{iopart}

\usepackage[british]{babel}
\usepackage{graphicx}
\usepackage{color}
\usepackage{iopams}
\usepackage{mathrsfs}
\usepackage[T1]{fontenc}
\usepackage{setstack}
\usepackage{bm}

\begin{document}

\title[Brownian motion and beyond]{Brownian motion and beyond: first-passage,
power spectrum, non-Gaussianity, and anomalous diffusion}
\author{Ralf Metzler}
\address{Institute of Physics and Astronomy, University of Potsdam,
14476 Potsdam-Golm, Germany}
\eads{\mailto{rmetzler@uni-potsdam.de}}

\begin{abstract}
Brownian motion is a ubiquitous physical phenomenon across the sciences.
After its discovery by Brown and intensive study since the first half of
the 20th century, many different aspects of Brownian motion and stochastic
processes in general have been addressed in Statistical Physics. In particular,
there now exist a very large range of applications of stochastic processes in
various disciplines. Here, we highlight some of the advances in stochastic
processes prompted by novel experimental methods such as superresolution
microscopy. Here we provide a summary of some of the recent developments
highlighting both the experimental findings and theoretical frameworks.
\end{abstract}

\section{Introduction}
\label{sec:Introduction}

In his seminal account Robert Brown reported the seemingly erratic motion of
granules of 1/4000th to 1/5000th of an inch in size, extracted from pollen
grains of the plant Clarkia pulchella. Being a botanist, Brown then went on
to prove that the observed jiggling motion is not due to the active motion
of "animalcules". Detecting the same type of motion of small granules of
inorganic matter---such as granite or a piece of the Egyptian Sphinx---he
demonstrated that the diffusive particle motion is in fact a true physical
phenomenon \cite{brown}. The precise understanding of Brownian motion was
then established in the groundbreaking works of Albert Einstein \cite{einstein},
William Sutherland \cite{sutherland}, Marian Smoluchowski \cite{smoluchowski},
and Pierre Langevin \cite{langevin}. Based on different arguments, they
derived the linear time dependence of the mean squared displacement (MSD)
\begin{equation}
\label{msd}
\langle\mathrm{r}^2(t)\rangle=\int\mathbf{r}^2P(\mathbf{r},t)dV=2dDt
\end{equation}
in $d$ spatial dimensions, and the Gaussian nature of the probability density
function (PDF) for a $\delta$-initial condition (Green's function)
\begin{equation}
\label{gauss}
P(\mathbf{r},t)=\frac{1}{(4\pi Dt)^{d/2}}\exp\left(-\frac{\mathbf{r}^2}{4Dt}
\right).
\end{equation}
This Gaussianity of $P(\mathbf{r},t)$ was independently established in the
discussion of the "random walk" in an exchange between Karl Pearson and John
William Strutt, third Lord Rayleigh \cite{pearson}.

The theoretical foundation of diffusion, especially the connection of the
diffusion coefficient $D$ with thermal energy $k_BT$ and thus Avogadro's
number $N_A$ in the Einstein-Smoluchowski-Sutherland relation
\begin{equation}
\label{einstein}
D=\frac{k_BT}{m\eta}=\frac{(R/N_A)T}{m\eta}
\end{equation}
in terms of the test particle mass $m$, the viscosity $\eta$ of the ambient
fluid, and the gas constant $R$, prompted a long series of ever improving
experiments on diffusive motion. Jean Perrin's systematic observations of
microscopic diffusing particles were groundbreaking in the introduction of
single particle tracking protocols \cite{perrin}. Additional noteworthy
contributions were due to Ivar Nordlund, who introduced time-resolved
recordings using a moving film plate and thus getting around the need to
average over---often not fully identical---ensembles of test particles
\cite{nordlund}, and Eugen Kappler, who studied the torsional diffusion
of a small mirror suspended on a long, thin quartz thread \cite{kappler}.
In fact, as this mirror experienced a restoring force by the twisting
thread---to first order, a Hookean force---it was Kappler who first
mapped out the Gaussian Boltzmann distribution of the equilibrium
distribution of the angles, to very high precision.

Today, following massive advances in microscopic techniques such as
superresolution microscopy (2014 Nobel Prize to Eric Betzig, W. E. Moerner,
and Stefan Hell) it is possible to follow submicron tracer particles or even
fluorescently labelled single molecules in highly complex environments such as
living biological cells \cite{orrit,lene_rev}. Thus, spatial resolutions in
the range of a few nanometres and time resolutions in the microsecond range
have been achieved \cite{rienzo}. Concurrently, supercomputing methods based
on molecular dynamics (2013 Nobel Prize to Arieh Warshel, Michael Levitt, and
Martin Karplus) have vastly improved, and huge data sets on complex systems
such as crowded lipid membranes are routinely produced \cite{bba_rev}.

The high spatiotemporal resolution of measured or simulated diffusive motion
in often complex environments have prompted numerous new developments in the
theoretical description of stochastic processes. Among these are new results
on Brownian first-passage processes, particularly, resolving the full density
of first-passage times. On a fundamental statistical physics level are questions
on ergodicity and reproducibility pinpointing whether the long time average of
single particle motion converges to the behaviour of an ensemble of identical
particles, or under what conditions repeated measurements can be expected to
deliver practically the same results. At finite measurement times, the time
averages of physical observables fluctuate from realisation to realisation,
and the quantification of these fluctuations allows one to extract information
on the system. Another feature observed in a growing number of experiments
is the non-Gaussianity of the PDF, and models to describe this behaviour are
called for. Finally, we mention anomalous diffusion, in which the MSD no
longer has the linear time dependence (\ref{msd}). In what follows we
highlight briefly these new developments. Section 2 addresses first-passage
processes beyond mean values, relevant in search and reaction processes. In
section 3 we discuss the quantification of single particle trajectories, in
particular, in terms of single trajectory power spectra. Section 4 addresses
the phenomenon of non-Gaussian diffusion and its theoretical description.
In section 5 we conclude and present a perspective.

\section{First-passage times: beyond mere means}

Consider a diffusing particle in one dimension that is initially released at
position $x_0>0$ at time $t=0$ on the half-line $x>0$ and with an absorbing
boundary at the origin. Solving the boundary value problem produces the
survival probability $\mathscr{S}(t)=\int_0^{\infty}P(x,t)dx$. Its negative
derivative is the first-passage time density \cite{redner}
\begin{equation}
\label{levysmirnov}
\wp(t)=-\frac{d\mathscr{S}(t)}{dt}=\frac{x_0}{(4\pi Dt)^{3/2}}\exp\left(-\frac{
x_0^2}{4Dt}\right),
\end{equation}
which is a one-sided probability density function of L{\'e}vy-Smirnov type. At
short times the first-passage time PDF has an exponential cutoff $\propto t^{-3/2}
\exp(-x_0^2/[4Dt])$, reflecting the fact that the particle needs a finite time to
move from $x_0$ to the origin. The long-time behaviour is given by the power-law
$\wp(t)\simeq t^{-3/2}$, causing the divergence of the mean first-passage time
$\langle t\rangle$. Note that this $3/2$ power-law asymptote based on Sparre
Anderson's result \cite{redner} is universal for
all Markovian symmetric random walks, in particular, with power-law jump length
distributions \cite{redner,koren}.
Once the first-passage process runs off in a finite domain---in higher dimensions,
with a finite-size target---the mean first-passage time $\langle t\rangle$ remains
finite. Remarkably, even for the mean first-passage time---or its "global" value,
averaged over all possible initial conditions---a number of novel, high-profile
results have been reported in recent years \cite{olivier}, underlining that the
mathematical development of first-passage processes is far from complete.

The point we want to stress here is that the situation becomes even more subtle
when we go beyond the (global) mean first-passage time and consider the full
PDF of first-passage times in a finite domain. A first indication that such a
study is relevant comes from the observation that in typical settings the times
of two independent realisations of first-passage events are disparate. In other
words, the distributions of the "uniformity index" $\omega=\frac{t_1}{t_1+t_2}$
of two first-passage times $t_1$ and $t_2$---with the same initial condition in
the same system---are effectively broad \cite{carlos}. On a more applied level,
for biochemical reactions in living cells, where the relevant molecules often
occur at minute, nanomolar concentrations, it is relevant to have information
on the diffusive reaction control beyond global mean first-passage times, and
the associated triggering reactions have a clear dependence on the initial
distance between the release of the particle and its designated binding spot
\cite{mirny}. As our discussion below shows, this property is beyond the concept
of the mean first-passage time. Indeed, distributions of first-passage times in this
context may indeed become broad \cite{maxscirep}. Experimentally, it is already
possible to resolve the production event of a single protein in a living cell
\cite{xie}, and it was shown that even relatively small, green fluorescent
proteins can be traced in live cells \cite{rienzo}. It will be possible to
follow individual biomolecules in their natural environment and determine
the molecule-resolved first-passage times.

\begin{figure}
\centering
\includegraphics[height=15.2cm,angle=270]{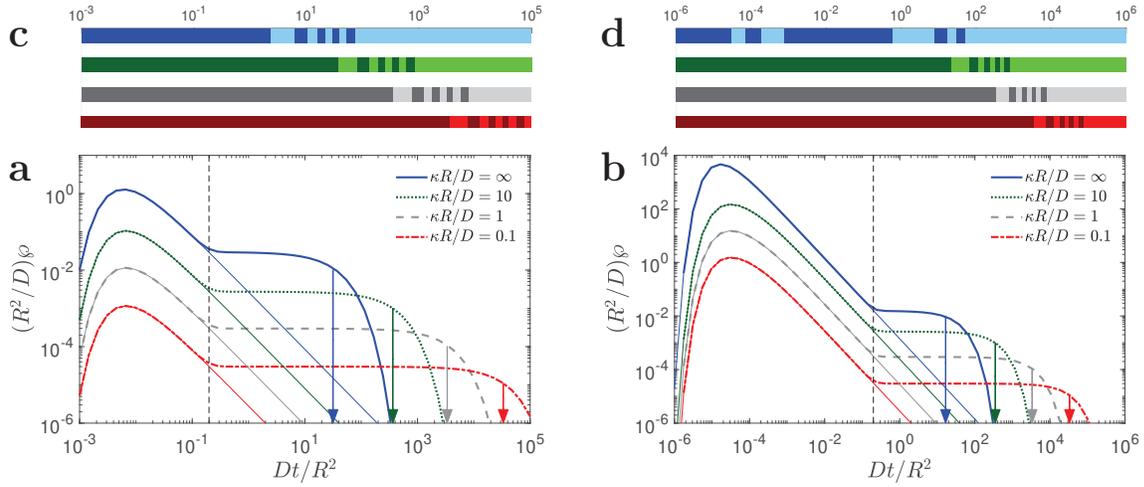}
\caption{Reaction time density $\wp(t)$ for a reaction on an inner target of
radius $\rho/R=0.01$, with starting point \textbf{a} $r/R=0.2$ and \textbf{b}
$r/R=0.02$ for four progressively decreasing (from top to bottom) values of the
dimensionless reactivity $\kappa R/D$ defined in the plot. The coloured
vertical arrows indicate the mean reaction times for these cases. The vertical
black dashed line indicates the crossover time $t_c=2(R−ρ)2/(D\pi^2)$ above
which the contribution of higher order eigenmodes become negligible. This
characteristic time marks the end of the hump-like region (L{\'e}vy-Smirnov
region specific to an unbounded system) and indicates the crossover to the
plateau region with equiprobable realisations of the reaction times. The
plateau region spans a considerable window of reaction times, especially
for lower reactivity values. Thin coloured lines show the reaction time
for the unbounded case ($R\to\infty$). Length and time units are fixed by
setting $R=1$ and $R^2/D=1$. Note the extremely broad range of relevant
reaction times (the horizontal axis) spanning over 12 orders of magnitude
for panel \textbf{b}. Coloured bar-codes \textbf{c} and \textbf{d} indicate
the cumulative depths corresponding to four considered values of $\kappa R/D$
in decreasing order from top to bottom. Each bar-code is split into ten regions
of alternating brightness, representing ten 10\%-quantiles of the distribution
(e.g., the first dark blue region of the top bar-code in panel \textbf{c}
indicates that 10\% of reaction events occur until $Dt=R^2\sim1$.) Figure
taken from \cite{denis1}.}
\label{denis}
\end{figure}

To quantify more precisely how broad the distribution of first-passage times
in a finite domain becomes, a Newton series technique for the calculations
of $\wp (t)$ for a range of different diffusion processes and spatial
dimensions was developed in hyperspherical domains \cite{aljaz}. In
all cases the PDF of first-passage times features distinct regions: (i)
an exponential suppression at very short times combined with a global
maximum, analogous to the behaviour captured in the L{\'e}vy-Smirnov type
PDF (\ref{levysmirnov}). The maximum value occurs at the point taken by
"direct" trajectories, that move relatively straight from the initial point
to the target. This is what may be called "geometry-control".  (ii) The
second region is given by a power-law decay with a process-dependent scaling
exponent. (iii) In the long time limit an exponential shoulder occurs whose
characteristic time depends both on the specific process and the details
of the diffusive domain. This exponential regime corresponds to "indirect"
trajectories in which the diffusing particles strays off its path to the
target and loses the memory to its initial condition due to collisions with
the outer boundary of the domain. Notably, the characteristic time encoded in
this exponential shoulder is closely related to the mean first-passage time.
Looking at the numbers this means that measuring the mean first-passage time
in such a process is in fact typically quite unlikely (see figure \ref{denis}).

A somewhat more general situation was analysed in \cite{denis,denis1},
namely, when we combine the diffusing particle with a target of finite
reactivity $\kappa$. For the mean reaction time, the contributions due to
the diffusive motion (characterised by the diffusivity $D$) and the reaction
(quantified by the reactivity $\kappa$), separate \cite{denis1},
\begin{equation}
\label{collins}
\langle t\rangle=\frac{(r-\rho)(2R^3-\rho r(r+\rho))}{6Dr\rho}+\frac{R^3-
\rho^3}{3\kappa\rho^2}
\end{equation}
for the case of a spherical domain with concentric target, where $R$ is the
outer (reflecting) radius, $\rho$ is the radius of the partially reactive
inner target, and $r$ is the radius of initial release of the diffusing
particle. The successful reaction in expression (\ref{collins}), which is
analogous to the famed Collins-Kimball rate \cite{collins}, thus requires
the diffusive search from the point of release to the target (the "diffusion
control", inversely proportional to $D$) plus the time to overcome the
"reaction barrier" (the "rate control", inversely proportional to $\kappa$).

The full first-passage time density to a successful reaction (or the
"reaction-time density") for this case is shown in figure \ref{denis}.
The features of the reaction-time density for a perfectly reactive target
($\kappa=\infty$) are identical to the above-described result for the
first-passage in \cite{aljaz}. Moreover, the initial exponential
suppression and the most likely first-reaction time, as well as the
intermediate power-law decay and the terminal exponential shoulder are
in fact commonly shared for finite reactivity $\kappa$, as well. However,
for decreasing reactivity the plateau region between the power-law and the
terminal exponential decay becomes increasingly pronounced. In addition to
the geometry-control in the most probable first-reaction time---whose value
is the same for all cases, as this value corresponds to direct trajectories
with immediate reaction success---the further defocusing of reaction-times at
finite $\kappa$ becomes more pronounced, the effect of "reaction-control"
\cite{denis1}. Depending on the details, individual
first-reaction times have a large probability to be shorter than the mean
first-reaction time, as shown in the "reaction depth" panels above the main
graph in figure \ref{denis}.

While in cases of macroscopic concentrations of reactive particles the mean
reaction time (or its inverse, the chemical rate constant) is a meaningful
quantity, at the minute concentrations in biochemical reactions, or in other
"first come first win" scenarios it loses its meaning, and a proper physical
understanding of the process requires cognisance of the full distribution of
first-passage or first-reaction times.

\section{Single trajectory mean squared displacement}

The most standard way to characterise a diffusion process is in terms of the MSD
(\ref{msd}). For anomalous, non-Brownian diffusion, the MSD is often of the
power-law form \cite{bouchaud,report,pccp}
\begin{equation}
\label{msd1}
\langle\mathbf{r}^2(t)\rangle\simeq K_{\alpha}t^{\alpha},
\end{equation}
where according to the value of the anomalous diffusion exponent $\alpha$ we
distinguish between subdiffusion ($0<\alpha<1$), Brownian diffusion ($\alpha
=1$), superdiffusion ($1<\alpha<2$), ballistic diffusion ($\alpha=2$), and
hyperdiffusion ($\alpha>2$). Examples for subdiffusion include the classical
charge carrier transport in amorphous semiconductors \cite{scher,schubert},
tracer diffusion in subsurface aquifers \cite{geo}, or the motion of passive
tracers in living biological cells \cite{cells}. Superdiffusion occurs in
weakly chaotic or fully turbulent flows \cite{turbulent}, or in actively driven
motion in cells \cite{cell_super}. We note that sometimes, also higher order
moments are being used, for instance, the skewness measuring the asymmetry of
a PDF involves the third order moment, and the kurtosis providing information
about the non-Gaussianity of a PDF is based on the fourth order moment,
see the next section on non-Gaussian distribution. Ratios of fourth order
moments versus the squared second order moment were shown to distinguish
different anomalous diffusion processes from another \cite{vincent}.

The MSD (\ref{msd}) or (\ref{msd1}) are ensemble quantities, based on the
evaluation of the second moment of the PDF $P(\mathbf{r},t)$. While this
ensemble-averaged MSD is a good quantity when a large ensemble of particles
are measured, in many modern setups such as single particle tracking,
typically relatively few individual particle trajectories $\mathbf{r}(t)$
of finite length $T$ are recorded. These are conventionally evaluated in
terms of the time-averaged MSD \cite{pccp,he,pt}
\begin{equation}
\label{tamsd}
\overline{\delta^2(t)}=\frac{1}{T-t}\int_0^{T-t}\Big[\mathbf{r}(t'+t)-
\mathbf{r}(t')\Big]^2dt'.
\end{equation}
For an ergodic process, $\overline{\delta^2(t)}$ in the limit of long
measurement times $T$ will converge to the MSD: $\lim_{T\to\infty}\overline{
\delta^2(t)}=\langle\mathbf{r}^2(t)\rangle$. This can be easily seen for a
Brownian process, which is self-averaging over sufficiently long times. Thus,
from a random walk perspective we say that the kernel in definition (\ref{tamsd})
is proportional to the number $n$ of jumps in the time interval $t$, $[\mathbf{r}
^2(t'+t)-\mathbf{r}^2(t')]^2\sim2dKt$, where $t=n\tau$ in terms of the number of
jumps in the interval $t$ and the time $\tau$ typically consumed for a single
jump. Then the diffusion coefficient can be identified as $K=\sigma^2/(2d\tau)$,
where $\sigma^2$ is the second moment of the jump length distribution
\cite{pccp,he,pt}. Evaluating the integral in (\ref{tamsd}) then immediately
produces that $\lim_{T\to\infty}\overline{\delta^2(t)}=2dKt$, proving ergodicity.
Note that for practical purposes in data analysis but also for calculations it
is useful to define the mean time-averaged MSD over a set of $N$ individual
trajectories denoted by the index $i$ \cite{he,pt},
\begin{equation}
\label{mtamsd}
\left<\overline{\delta^2(t)}\right>=\frac{1}{N}\sum_{n=1}^N\overline{\delta^2_i(t)}.
\end{equation}

Anomalous diffusion is non-universal: there exist many different stochastic
processes giving rise to the power-law form (\ref{msd1}) of the MSD. The best
known examples include continuous time random walks whose sojourn (trapping)
times $\tau$ are power-law distributed, $\psi(\tau)\simeq\tau_0^{\alpha}/\tau^
{1+\alpha}$, such that the characteristic sojourn time diverges \cite{scher}.
Due to the lack of time scale, individual sojourn times may occur whose length
is of the order of the duration of the process, no matter how long this process
evolves. As a consequence, the time-averaged MSD behaves fundamentally differently
from the ensemble MSD. Thus, while the MSD is of power-law form (\ref{msd1}), the
time-averaged MSD $\overline{\delta^2(t)}$ always remains a random quantity, even
in the long measurement time limit \cite{pccp,he,pt}. Its mean can be shown to
follow the relation ($t\ll T$) \cite{he,pt}
\begin{equation}
\label{tamsd_ctrw}
\left<\overline{\delta^2(t)}\right>\sim2dK_{\alpha}\frac{t}{T^{1-\alpha}}.
\end{equation}
The lag time ($t$) dependence is thus linear, as if the process were Brownian.
The anomaly of the process can only be seen in the explicit dependence on the
observation time $T$: the longer the process evolves, the more it slows down,
reflecting the occurrence of ever longer waiting times, on average. This
non-stationary behaviour is called ageing and is measured experimentally
\cite{weigel}.

Ageing may also be relevant in a somewhat different setting: start a continuous
time random walk process (or another ageing stochastic process) at some initial
time, but commence the actual measurement at $t_a>0$. When we count time $t$
from this moment, the ageing time-averaged MSD reads \cite{johannes}
\begin{equation}
\overline{\delta^2_a(t)}=\frac{1}{T-t}\int_{t_a}^{T+t_a-t}\Big[\mathbf{r}(t'+t)
-\mathbf{r}(t')\Big]^2dt'
\end{equation}
For a continuous time random walk process with scale-free waiting times, the
resulting behaviour is \cite{johannes}
\begin{equation}
\label{atamsd}
\left<\overline{\delta^2_a(t)}\right>\simeq\Lambda_{\alpha}(t_a/T)\left<
\overline{\delta^2(t)}\right>,\,\,\,\Lambda_a(z)=(1+z)^{\alpha}-z^{\alpha}.
\end{equation}
Here the purely multiplicative factor $\Lambda_{\alpha}$ depends on the ratio
$t_a/T$ of the two time scales of the system, while on the right hand side of
expression (\ref{atamsd}) the non-aged time-averaged MSD (\ref{tamsd}) appears.
In the time average, that is, the ageing time $t_a$ enters in a simpler way
than in the corresponding ageing MSD, that exhibits a crossover behaviour from
$\langle\mathbf{r}^2_a(t)\rangle\simeq t/t_a^{1-\alpha}$ for $t_a\gg t$ to
$\simeq t^{\alpha}$ for $t\gg t_a$ \cite{johannes,eli}.

Non-ergodicity and ageing not only occur in scale-free continuous time random
walks. While diffusion on the infinite cluster of a critical percolation
network is ergodic, when we consider all clusters of a percolation network,
due to the random seeding of the walkers on clusters of various sizes,
non-ergodic behaviour results \cite{youssof}. Other examples, in which the
behaviours with respect to non-ergodicity and ageing have analogous expressions
as relations (\ref{msd1}), (\ref{tamsd_ctrw}), and (\ref{atamsd}) are scaled
Brownian motion defined in terms of a Markovian Langevin equation with
time-dependent diffusion coefficient, $K(t)\simeq t^{\alpha-1}$ \cite{sbm} as
well as heterogeneous diffusion processes with position-dependent diffusivity
$K(x)=K_0|x|^{2-2/\alpha}$ \cite{hdp}. Ageing also affects other quantities of
the associated process. For instance, it may induce a population splitting into
mobile and immobile subpopulations: for scale-free continuous time random walks
the probability that at least one jump occurs during the measurement period
$T$ decays with the ageing time $t_a$ as $m_{\alpha}\simeq(T/t_a)^{1-\alpha}$
due to the increasing probability of ever larger sojourn times with growing
$t_a$ \cite{johannes}. Similar effects occur in heterogeneous diffusion
processes \cite{andreypccp}. Another quantity that is directly affected by
ageing is the first-passage time density \cite{schubert, henning}. Conversely,
ergodic anomalous diffusion processes exist in the form of processes driven
by stationary but long-range correlated fractional Gaussian noise, namely,
fractional Brownian motion \cite{mandelbrot,deng} and fractional Langevin
equation motion \cite{deng,goychuk}.

For finite-time measurements even a Brownian process will lead to non-identical
results from one to the next trajectory. In the plot of the time-averaged MSD
this effect will produce certain amplitude fluctuations, at a given lag time,
between individual $\overline{\delta^2_i(t)}$. For scale-free continuous time
random walk processes, due to the much more likely occurrence of extreme sojourn
times, the amplitude fluctuations will be considerably more pronounced. Such
amplitude scatter can be quantified in terms of the distribution $\phi(\xi)$,
where the dimensionless variable $\xi=\overline{\delta^2(t)}/\langle\overline{
\delta^2(t)}\rangle$ measures how much the time-averaged MSD $\overline{\delta
^2(t)}$ deviates from the mean $\langle\overline{\delta^2(t)}\rangle$ \cite{he,
pt}. The amplitude scatter distribution $\phi(\xi)$ and its variance, the
ergodicity breaking parameter $\mathrm{EB}=\langle\xi^2\rangle-1$ have been
calculated for a variety of processes \cite{pccp,he,sbm,hdp}, and they encode
distinct behaviours as functions of lag and measurement time for the different
processes \cite{pccp}.

\section{Single trajectory power spectra}

The MSD and time-averaged MSD are standard measures for stochastic processes.
However, there exists an alternative approach to quantify a diffusive dynamics,
especially used in experimental data analysis, the power spectrum.

The standard textbook setting for spectral analyses is based on the so-called
power spectral density (PSD) $\mu(f)$. To that end, the PSD is obtained by
first calculating the Fourier transform  of the individual trajectory $X(t)$,
measured over over the finite observation time $T$,
\begin{equation}
\label{singlepow}
S(f,T)=\frac{1}{T}\left|\int_0^Te^{ift}X(t)dt\right|^2.
\label{PSD1}
\end{equation}
Here $f$ denotes the frequency. The single-trajectory quantity $S(f,T)$ for
finite observation times $T$ naturally is a random variable, similar to our
discussion of the time-averaged MSD above. The standard PSD is obtained from
$S(f,T)$ as the ensemble average over all possible trajectories. With the
additional long-measurement-time limit $T\to\infty$, the standard PSD yields in
the form \cite{norton,PSD_BM,PSD_fBM}
\begin{eqnarray}
\label{main}
\nonumber
\mu(f)&=&\lim_{T\to\infty}\frac{1}{T}\left\langle\left|\int_0^Te^{ift}X(t)dt\right|^2
\right\rangle\\
&=&\lim_{T\to\infty}\frac{1}{T}\int_0^T\int_0^T\cos(f[t_1-t_2])
\langle X(t_1)X(t_2)\rangle dt_1dt_2,
\end{eqnarray}
where angular brackets represent the statistical averaging and $\langle X(t_1)
X(t_2)\rangle$ is the autocorrelation function of the process $X(t)$.

For the vastly growing number of single particle tracking experiments, the
typical situation is that relatively few individual trajectories are garnered
with a finite observation time $T$. Thus both the statistical averaging and
the long time limit entering the definition (\ref{main}) are problematic. As
an alternative, practicable approach we therefore recently defined the single
trajectory power spectral analysis based on expression (\ref{singlepow})
\cite{PSD_BM}. In general, the single-trajectory PSD (\ref{PSD1}) will not only
be a function of the frequency $f$ but also of the observation time $T$. In
addition, fluctuations of $S(f,T)$ between results for individual trajectories
will occur, even for normal Brownian motion \cite{PSD_BM}---in analogy to the
amplitude fluctuations of the time-averaged MSD discussed above. While such
trajectory-to-trajectory fluctuations may be mitigated by taking statistical
averaging, we argue that important information may be drawn from these
fluctuations---similar to the information from the amplitude scatter
distribution $\phi(\xi)$ on the time-averaged MSD above.

\begin{figure}
\centering
\includegraphics[width=15.8cm]{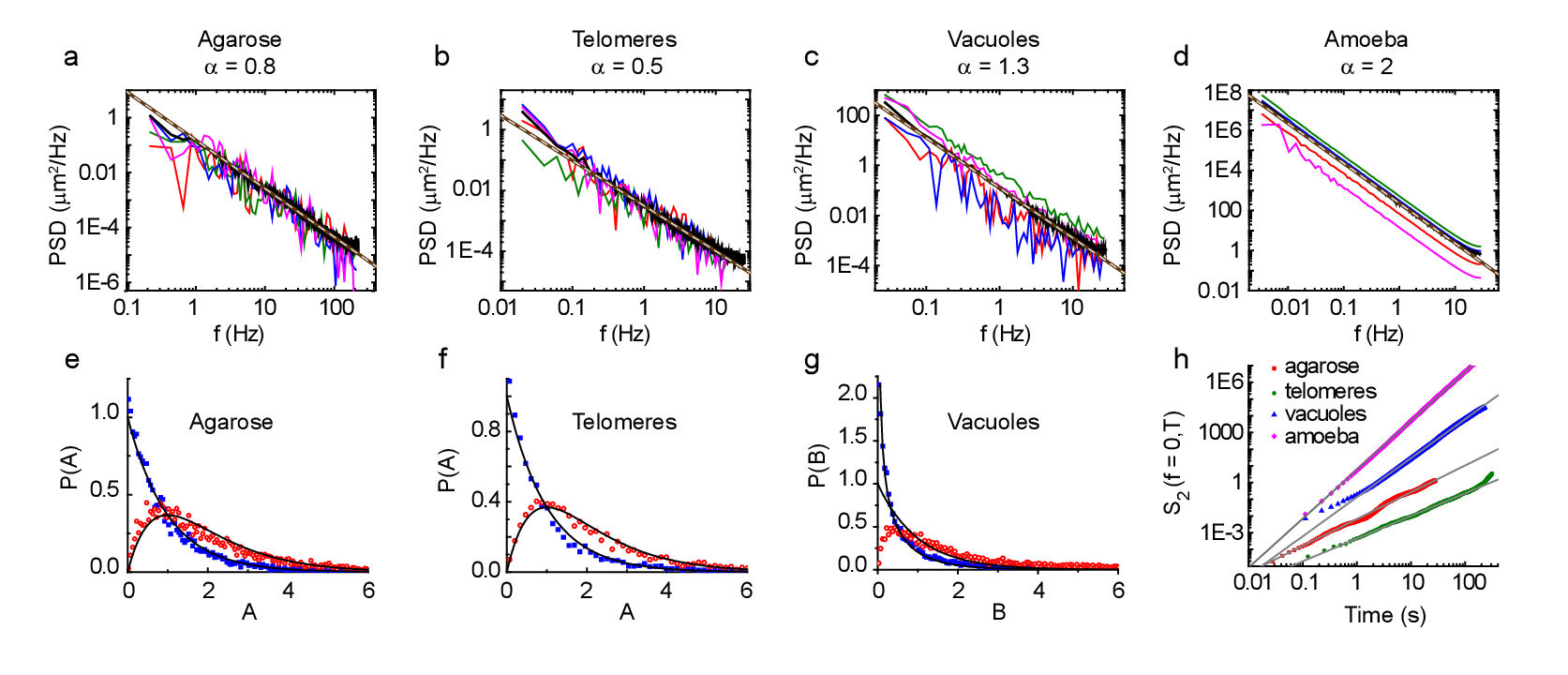}
\caption{Power spectral analysis of experimental data sets, taken from
\cite{PSD_fBM}. (a-d) Single-trajectory PSD of representative trajectories
along with the ensemble-averaged PSD for telomeres in the nucleus of HeLa
cells, 50-nm nanoparticles in 1.5\% agarose gel, intracellular vacuoles
within amoeba, and the motion of amoeba. The anomalous diffusion exponent
in the panels indicates sub- and superdiffusive dynamics. The dashed thick
lines show the power-laws $1/f^{1.49}$ for panel (a), $1/f^{1.76}$ for panel
(b) and $1/f^2$ for panels (c) and (d). In each case, the PSDs of four
trajectories are presented (log-sampled with a factor 1.1 for clarity)
together with the ensemble-averaged PSD (thicker black lines, $n=19, 20, 50,$
and $4$ trajectories for telomeres, nanoparticles, vacuoles, and amoeba,
respectively). (e-g) Amplitude distribution of the PSD for one and two
dimensions. (h) Zero frequency PSD, showing the ageing dependence mentioned
in the text. For more details see \cite{PSD_fBM}.}
\label{fig_power}
\end{figure}

The single trajectory power spectrum has so far been analysed for Brownian
motion, fractional Brownian motion, and scaled Brownian motion \cite{PSD_BM,
PSD_fBM,power2}. In all these results the single-trajectory PSD (\ref{PSD1})
is proportional to the ensemble-averaged PSD (\ref{main}), where depending
on the process parameters the scaling with frequency exhibits the Brownian
like $\simeq f^{-2}$ behaviour or a scaling exponent depending explicitly
on the anomalous diffusion exponent $\alpha$. Depending on the exact process
the single-trajectory PSD (\ref{PSD1}) may feature an explicit, ageing dependence
on the observation time $T$. Naturally, see the discussion of the time-averaged
MSD above, individual finite-$T$ realisations will differ from each other by a
random numerical factor in the single-trajectory PSD (\ref{PSD1}). The
distribution of this amplitude was calculated analytically for Brownian motion
and fractional Brownian motion \cite{PSD_BM,PSD_fBM}. The shape of this
distribution depends on whether one analyses the full three-dimensional motion,
its two-dimensional projection typically measured by single particle tracking
experiments, or the projection onto one dimension. Figure \ref{fig_power}
demonstrates how well the experimentally observed behaviour matches the
analytically predicted behaviour for four different systems in both
subdiffusive and superdiffusive domains.

A special r{\^o}le is played by the coefficient of variation $\gamma$ in the case
of fractional Brownian motion, that can be described in terms of an overdamped
Langevin equation driven by Gaussian but power-law correlated noise with noise
autocorrelation $\xi(t)\xi(t')\sim\alpha(\alpha-1)K_{\alpha}|t-t'|^{\alpha-2}$
\cite{deng,pccp}. For subdiffusion, the coefficient is negative, reflecting the
antipersistent nature of the process, while for superdiffusion persistence is
observed. As function of $\omega=fT$, $\gamma$ has the unique value $\sqrt{2}$
independent of the anomalous diffusion exponent $\alpha$. In the large-$\omega$
limit, the result for $\gamma$ reads
\begin{equation}
\label{gammas}
\gamma\sim\left[1+\left(1+c_{\alpha}\omega^{1-\alpha}\right)^{-2}\right]^{1/2}, 
\end{equation}
where $c_{\alpha}=\Gamma(1+\alpha)\sin(\pi\alpha/2)$. Remarkably, when we take
$\omega\to\infty$ there only exist three different values of $\gamma$: $\sqrt{2}$
results for superdiffusion ($\alpha>1$), $\sqrt{5}/2$ is obtained for normal
Brownian motion ($\alpha=1$), and $1$ is the limiting value for subdiffusion
($\alpha<1$) \cite{PSD_fBM}. As shown from relatively few and short trajectories,
$\gamma$ allows one to distinguish the three regimes (superdiffusion, normal,
diffusion, subdiffusion) even significantly away from the limit $\omega\to
\infty$: it is sufficient to see whether the values depart from the Brownian
one to show a tendency of growth or decrease \cite{PSD_fBM}.

Commonalities and differences between these Gaussian processes need careful
analysis. In many respects the behaviour is the same for different processes
when we look at one observable, for instance, the scaling exponent of the
single-trajectory PSD, while the ageing dependence is different, etc. For
the future a complete analysis of the single-trajectory PSD behaviour of
a more exhaustive range of anomalous stochastic processes is called for,
ultimately providing a very powerful tool to analyse measured time series,
as an alternative to moment-based analyses.

\section{Non-Gaussian diffusion processes: normal and anomalous}

The central limit theorem is a cornerstone of statistical physics and
mathematical statistics: according to this theorem the possible values
of a properly scaled sum of independent, identically distributed (IID)
random variables are governed by the normal Gaussian PDF in the limit of a
large number of entries \cite{feller}.  This convergence is strong in the
sense that it is independent of the exact form of the distribution of the
component IID variables, if only they are characterised by a finite variance.
An outstanding example for this mathematical statement is the convergence of
the position distribution of a normal random walk to the Gaussian (\ref{gauss})
\cite{vankampen}. In particular, this solution (\ref{gauss}) of the diffusion
equation features the scaling variable $|\mathbf{r}^2|/t$, giving rise to
the Brownian (Fickian) MSD (\ref{msd}) with its linear time dependence.

Often, the observation of an MSD of the linear form (\ref{msd}) is taken to
imply that we are dealing with normal Brownian motion, and that the PDF of
the process therefore has to be the Gaussian (\ref{gauss}). A number of recent
data from a range of different systems demonstrate, however, that a linear
MSD (\ref{msd}) can come along with highly non-Gaussian forms of the PDF $P(
\mathbf{r},t)$ \cite{Wang:BYNG2,Wang:BYNG3}. For instance, the motion of
biomacromolecules, proteins and viruses along lipid tubes and through actin
networks \cite{Wang:BYNG2,Wang:BYNG3}, as well as along membranes and inside
colloidal suspension \cite{Goldstein:BYNG9} and colloidal nanoparticles
adsorbed at fluid interfaces \cite{xue:BYNG8,wang,Dutta-Chak} show this
"Brownian yet non-Gaussian" behaviour. A similar combination of the law
(\ref{msd}) with non-Gaussian properties was observed in ecological processes
of organism movement and dispersal \cite{Hapca:BYNG4,beta}. There exist also
processes, that are Brownian but non-Gaussian in certain time windows of
their dynamics observed for the dynamics of disordered solids, such as glasses
and supercooled liquids \cite{Kob1,Kob:BYNG5,Sciortino} as well as for
interfacial dynamics \cite{samanta:BYNG7,skaug}.

To see how a non-Gaussian PDF may arise while the MSD is linear in time,
consider a mixture of diffusing particles with non-identical diffusivities
$D$. An example could be commercial tracer beads that always have a certain
size distribution due to imperfections in the manufacturing process. Indeed,
already Jean Perrin faced this problem in his early single particle tracking
experiments. While each particle is Brownian and, for its own $D$ value,
characterised by a Gaussian $P(\mathbf{r},t|D)$\footnote{We use the explicit
conditional probability notation to indicate the $D$-dependence of
the PDF.}, if we measure the PDF for the entire "ensemble" of the non-identical
particles the result will be the average
\begin{equation}
\label{superstat}
P(\mathbf{r},t)=\int_0^{\infty}p(D)P(\mathbf{r},t|D)dD.
\end{equation}
Here $p(D)$ quantifies the distribution of diffusivities among the tracer
particles. In fact, the formulation (\ref{superstat}) is identical to the
concept of superstatistics formulated by Beck and Cohen \cite{Super1}, see
also \cite{Superst}. Their original scenario for relation (\ref{superstat})
was that individual particles move in different regions characterised by
different $D$. In this scenario, of course, each particle will eventually
reach the border of its seed region and move to a region with a different
$D$, and $p(D)$ would become explicitly time dependent. However, in the
superstatistical formulation (\ref{superstat}) $p(D)$ is time independent.
We note that the superstatistical formulation was also achieved starting
from a stochastic Langevin equation \cite{straeten}. Moreover, a similar,
random-parameter formulation of diffusion processes is given by the
concept of (generalised) grey Brownian motion \cite{ggbm,vittoria}.

The first results in \cite{Wang:BYNG2,Wang:BYNG3} of the non-Gaussian
distribution $P(\mathbf{r},t)$ were the exponential or "Laplace" distribution.
One can show \cite{prx} that this form of $P(\mathbf{r},t)$ uniquely emerges
from an exponential distribution $p(D)$. More complicated forms of $p(D)$ are
often found in terms of generalised gamma distributions, as observed in
\cite{Hapca:BYNG4}, or stretched Gaussian shapes \cite{beta}. Superstatistical
and ggBm formulations based on the generalised gamma distribution were introduced
in \cite{Hapca:BYNG4,vittoria,gengam}.

In its formulation above superstatistics cannot account for the crossover to
a Gaussian PDF at times longer than some correlation time observed in some of
the experiments \cite{Wang:BYNG2,Wang:BYNG3}. This was achieved by Chubinsky
and Slater in their model of "diffusing diffusivity" \cite{gary}. This approach
was further developed by Jain and Sebastian \cite{klsebastian},
Chechkin \etal \cite{prx}, Tyagi and Cherayil \cite{tyagi},
Lanoisel{\'e}e and Grebenkov \cite{Lanoiselee}, as well as Sposini \etal
\cite{vittoria}. The basic idea by Chubinsky and Slater is that the diffusion
coefficient in a single trajectory is a stochastic quantity, changing its
value perpetually along the trajectory of the tracer particle. Physically,
this is a simplified picture for a particle moving in a heterogeneous
environment, imposing continuous changes in the particle mobility along
its path. Concretely, in a minimal formulation of the diffusing diffusivity
model, this motion can be captured by the set of coupled stochastic equations
\cite{prx}
\numparts
\label{langmin}
\begin{eqnarray}
\label{lang1}
\frac{d}{dt}\mathbf{r}(t)&=&\sqrt{2D(t)}\bm{\xi}(t),\\
\label{lang2}
D(t)&=&\mathbf{Y}^2(t),\\
\label{lang3}
\frac{d}{dt}\mathbf{Y}(t)&=&-\frac{1}{\tau}\mathbf{Y}+\sigma\bm{\eta}(t).
\end{eqnarray}
\endnumparts
Here expression (\ref{lang1}) is the Langevin equation for a a particle
driven by the white Gaussian noise $\bm{\xi}(t)$. However, the associated
amplitude contains the explicitly time-dependent diffusion
coefficient. This property is specified by
equations (\ref{lang2}), that maps $D$ onto the squared auxiliary quantity
$\mathbf{Y}$ thus guaranteeing positivity of the diffusivity, and (\ref{lang3}). 
The latter, stochastic equation describes the time evolution of $\mathbf{Y}$
driven by another white Gaussian noise $\bm{\eta}(t)$. However, in contrast
to equation (\ref{lang1}), the motion of $\mathbf{Y}$ is confined and thus
will relax to equilibrium above the crossover time $\tau$. In fact, equation
(\ref{lang3}) is the famed Ornstein-Uhlenbeck process \cite{vankampen}. In the
analysis of \cite{prx} it was shown that this formulation of the
diffusing diffusivity model at short times reproduces the superstatistical
approach, while at times longer than the correlation time $\tau$ of the
auxiliary $\mathbf{Y}$ process a crossover occurs to a Gaussian PDF
characterised by a single, effective diffusion coefficient. This crossover
can be conveniently characterised by the kurtosis $K=\langle\mathbf{r}^4(t)
\rangle/\langle\mathbf{r}^2(t)\rangle^2$ \cite{prx}. More technically,
the formulation in terms of the minimal model (\ref{lang1}) to (\ref{lang3})
corresponds to a subordination approach, which is helpful in obtaining exact
analytical results and in formulating a
two-variable Fokker-Planck equation for the diffusing diffusivity process
\cite{prx}. We note that the first-passage behaviour of the
diffusing diffusivity model was analysed in \cite{grebenkov,vittoria1}.

\begin{figure}
\centering
\includegraphics[width=5.12cm]{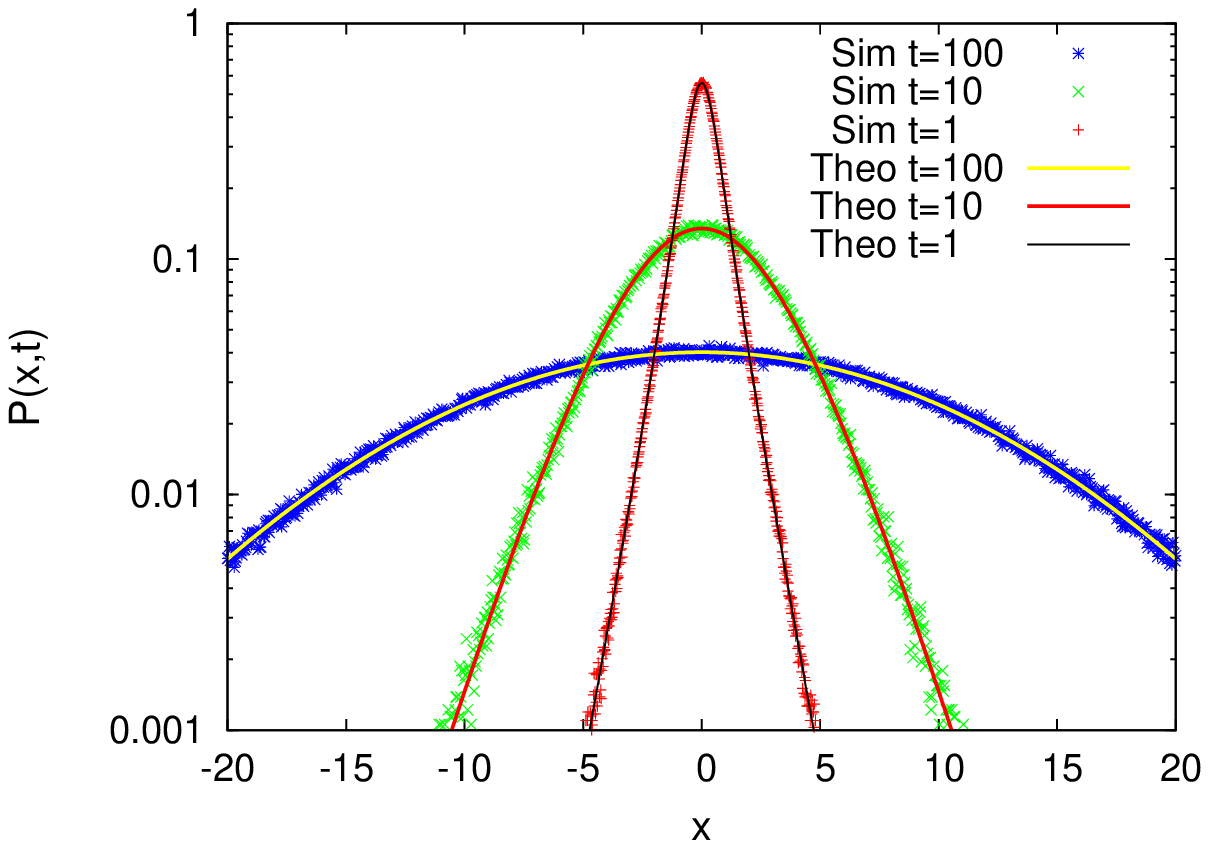}
\includegraphics[width=5.12cm]{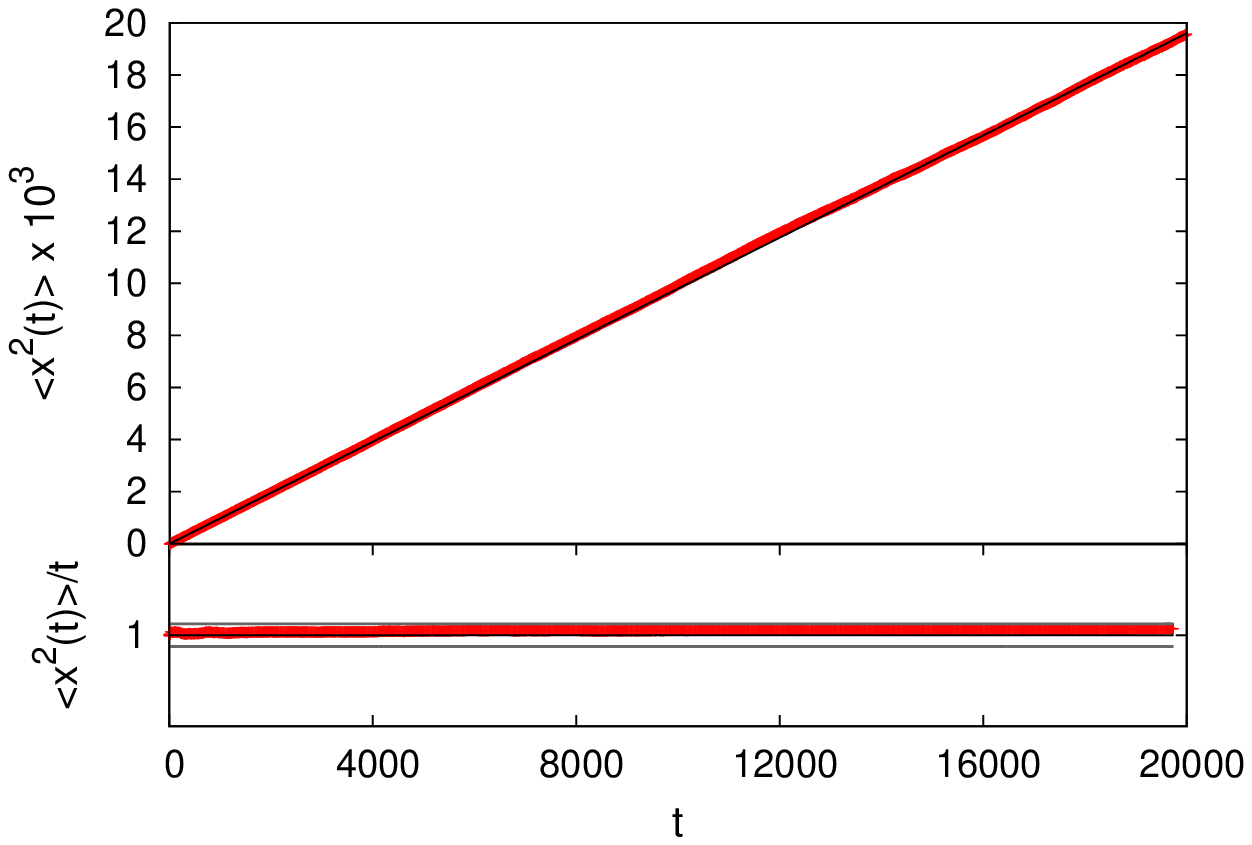}
\includegraphics[width=5.12cm]{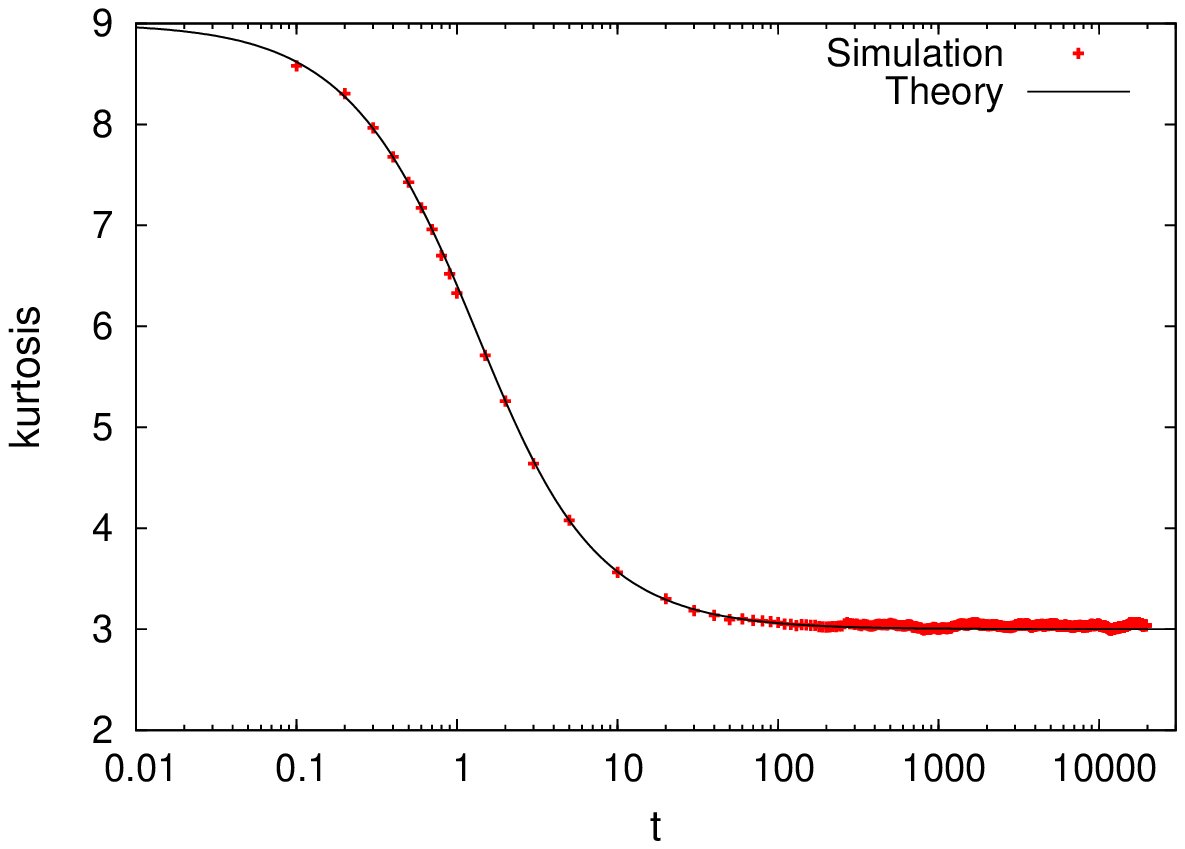}
\caption{Behaviour of the minimal model for diffusing diffusivity, equations
(\ref{lang1}) to (\ref{lang3}) in the one-dimensional case, figures reproduced
from \cite{prx}. Left: PDF $P(x,t)$ at different times, demonstrating
the crossover from the short-time exponential to the long-time Gaussian form,
shown here for simulations and the theoretical result.
Middle: the MSD shows a linear behaviour with constant coefficient, as seen in
the lower panel, in which the MSD/$t$ is shown.
Right: the kurtosis crosses over from the value $K=9$ for a one-dimensional
Laplace distribution to the value $K=3$ for a one-dimensional Gaussian; the
crossover time corresponds to the preset value $\tau=1$.}
\label{diffdiff}
\end{figure}

In figure \ref{diffdiff} we show the behaviour encoded in the minimal
diffusing diffusivity model (\ref{lang1}) to (\ref{lang3}). The three panels
respectively show the crossover from an initial Laplace distribution with
exponential tails (Left), the fact that the MSD of the process always is
linear in time with a constant coefficient (Middle), and the crossover
behaviour measured by the kurtosis (Right). This behaviour is characteristic
for an equilibrium condition of the auxiliary variable $\mathbf{Y}$. The
more general situation for a non-equilibrium initial condition with crossovers
in the associated MSD is analysed in \cite{vittoria}. We remark that the
diffusing diffusivity model developed here is closely related to the
Cox-Ingersoll-Ross (CIR) model for monetary returns which is widely used
in financial mathematics \cite{cir}.

What about anomalous diffusion processes? Fractional Brownian motion (FBM) and
fractional Langevin equation (FLE) motion\footnote{The FLE is a version of the
generalised Langevin equation with a power-law friction kernel \cite{goychuk,
lutz}.} are both processes driven by power-law correlated (fractional) Gaussian
noise, and are therefore characterised by a Gaussian PDF. For the motion of
constituent molecules in membrane systems it was shown in a supercomputing
study that the dynamics is Gaussian and driven by fractional Gaussian noise
\cite{prl}, however, when the membrane is crowded by large embedded proteins,
strongly non-Gaussian behaviour occurs with intermittent diffusivity occurs
\cite{ilpo_prx}. In single particle tracking experiments in heterogeneous
membranes an exponential distribution of the diffusivity were shown along
with a Laplace distribution of the PDF \cite{natcom}. Finally we mention the
study \cite{spako} of tracer diffusion in bacteria and yeast cells, where
a Laplace-PDF and exponential diffusivity distribution were presented and
motivated by a superstatistical FBM approach. Similarly to this approach,
a superstatistical generalised Langevin equation model was studied by Beck
and van der Straeten \cite{straeten1}, while a more general approach for a
superstatistical generalised Langevin equation was introduced by {\'S}l\k{e}zak
\etal \cite{jakub} in which it was shown that the distribution of the position
variable is characterised by a relaxation from a Gaussian to a non-Gaussian
distribution.

Random parameter diffusion models are very actively studied, and we can
here only give a limited overview. Apart from the developments sketched
above we mention the study by Cherstvy \etal \cite{andrey_pccp} in which
scaled Brownian motion for massive and massless particles was analysed
for a Rayleigh distribution of the diffusion coefficient. Stylianidou \etal
\cite{spako_pre} show that in a random barrier model anomalous diffusion with
exponential-like step size distribution and anticorrelations emerge, similar
to the behaviour measured by Lampo \etal \cite{spako}, with a crossover to
Brownian and Gaussian behaviour at sufficiently long times. Sokolov \etal
compare the diffusing diffusivity model with the emerging dynamics when the
quenched nature of a disordered environment is explicitly taken into account
\cite{igor_quenched}. Moreover, we mention a study by Barkai and Burov
\cite{staseli}, in which the authors use extreme value statistic arguments
to derive a robust exponential shape of the displacement PDF. Finally, in a
recent work {\'S}l\k{e}zak \etal \cite{jakub2} show that random coefficient
autoregressive processes of the ARMA type can be used to describe Brownian
yet non-Gaussian processes, and thus connect the world of physics of such
dynamics with the world of time series analysis.

We close this section with the remark that continuous time random walk
processes with scale-free waiting times inherently have a non-Gaussian
distribution \cite{pccp,report}, as does diffusion on fractal supports
such as percolation clusters close to criticality \cite{havlin}. Finally,
a completely different mechanism for non-Gaussianity is currently being
explored. Namely, while normal Brownian motion initiated right next to a
reflecting boundary will develop as a half-Gaussian, fractional Brownian
motion with its correlated increments shows pronounced deviations from
the Gaussian form in the vicinity of the boundary, and does also not
converge to a constant distribution in a finite box domain \cite{newfbm}.

\section{Conclusions}

Despite its relatively long history and the existence of numerous textbooks
the theory of Brownian motion is still far from complete. Less surprisingly,
anomalous stochastic processes are still actively studied. We here mentioned
some of the recent developments, including the calculation of the full
distributions of first-passage times in generic confined volumes and for
targets with finite reactivity, the theory of time-averaged moments and the
feature of non-ergodicity and ageing, the single-trajectory power spectral
density, and the emergence of non-Gaussian distributions in heterogeneous
media. These developments are motivated by novel experimental techniques,
for instance, superresolution microscopy and/or single particle tracking
in complex environments such as living biological cells. In this endeavour,
however, theory also feeds back to experiment. A prime example is the analysis
of Perrin that would not have been possible without the theories by Einstein
and Smoluchowski. The recent mathematical results for stochastic processes 
presented here will allow experimentalists to compare their observations
with these predictions and help them to extract the physical parameters
and identify the underlying stochastic mechanisms.

One of the major lessons coming from the current theoretical analysis of
stochastic processes is that, instead of aiming at producing smooth curves
in terms of averaging over as many particles as possible, valuable information
can indeed be gained from the fluctuations of the measured quantities. Thus,
for the amplitude scatter distribution $\phi(\xi)$ of the time-averaged MSD
distinct patterns emerge for different stochastic processes, helping in
distinguishing these physical mechanisms when analysing data. Similarly the
lag and observation time dependence of the ergodicity breaking parameter
$\mathrm{EB}$ has a similar diagnostic r{\^o}le. Amplitude fluctuations have
also been calculated for the single-trajectory power spectral density. Once
this property is known for a larger class of stochastic processes, these
fluctuations will similarly act as a criterion for model selection. To
acknowledge the presence of strong fluctuations is also important given
our discussion of first-passage times and their defocusing.

Data analysis is becoming ever more relevant as more and higher quality
data are being obtained. As we saw, there exist numerous, qualitatively
different anomalous stochastic processes. To learn about the physics of
a system, the exact underlying stochastic mechanism needs to be identified,
along with reliable values for the systems parameters. This is currently
being investigated, using different approaches. We here mention Bayesian
based maximum likelihood systems tailored for diffusive systems \cite{bayes},
as well as machine learning suites \cite{yaelgorka}. Considerable advances
in this field over the coming years are to be expected. In parallel,
theorists are developing new tools for the data analysis, such as the
moment or power spectral analysis mentioned here, or other methods such
as the p-variation technique \cite{pvar}, apparent diffusivity distributions
\cite{radons}, covariance-based estimators \cite{henrik}, or the codifference,
that is able to detect ergodicity breaking and non-Gaussianity in measured
data \cite{jakub1}.

The exploration of stochastic processes is still going strong, new and
relevant theoretical results allowing experimentalists to focus their
studies, while concurrently new types of experiments and ever improving
precision, resolution, and sheer amounts of measured data pose new
challenges for the theoretical analysis.

\ack RM acknowledges funding from Deutsche Forschungsgemeinschaft (DFG
code ME 1535/7-1) as well as from the Foundation for Polish Science
(Fundacja na rzecz Nauki Polskiej) within an Alexander von Humboldt
Polish Honorary Research Scholarship.

\end{document}